\documentclass[aps,prc,twocolumn,showpacs]{revtex4-1}
\usepackage{graphicx}
\usepackage{dcolumn}
\usepackage{bm}

\begin{document}

\title{Uncertainties in the analysis of neutron resonance data}

\author{J. F. Shriner, Jr.$^a$, H. A. Weidenm\"{u}ller$^b$, and G. E. Mitchell$^c$}
\email{jshriner@tntech.edu, Hans.Weidenmueller@mpi-hd.mpg.de, mitchell@tunl.duke.edu}
\affiliation{$^a$Department of Physics, Tennessee Technological University, Cookeville, TN 38505, USA \\ $^b$Max-Planck-Institut f{\"u}r Kernphysik, 69029 Heidelberg, Germany \\ $^c$North Carolina State University, Raleigh, North Carolina 27695, USA and \\ Triangle Universities Nuclear Laboratory, Durham, North Carolina 27708, USA}

\begin{abstract}

Recent analyses of the distribution of reduced neutron widths in the
Nuclear Data Ensemble (NDE) and in the Pt isotopes find strong
disagreement with predictions of random-matrix theory. These analyses
combine the maximum-likelihood method with a cutoff on the reduced
neutron widths. We show that the method introduces a systematic error
(the ``cutoff error''). That error (seemingly taken into account for
the Pt data) increases with increasing cutoff and decreasing number of
data points. We reanalyse the NDE taking the cutoff error into
account. While differing in detail, our results confirm the earlier
conclusion that the NDE disagrees significantly from RMT predictions.

\end{abstract}


\maketitle


\section{Purpose}

Early work~\cite{Haq82, Boh83, Boh85} on the distribution of widths and
spacings of neutron resonances in a set of nuclei (the ``Nuclear Data
Ensemble'') (NDE) indicated agreement with predictions of Random
Matrix Theory (RMT). Firm and unbiased conclusions can only be drawn,
however, if $s$-wave and $p$-wave neutron resonances are cleanly
separated. That was not possible or not done at the time. The problem
was emphasized in Ref.~\cite{Koe11}, and in Refs.~\cite{Koe10, Koe11}
it was addressed with the help of a cutoff on the reduced neutron
widths that depends on neutron resonance energy, variation of a
technique originally applied in Ref.~\cite{Cam94} to a much smaller
data set. The ensuing re-analysis of the NDE~\cite{Koe11} and the
analysis of new data on the Pt isotopes~\cite{Koe10} using different
versions of the cutoff method have both cast serious doubt on the
validity of random-matrix theory (RMT) in nuclei. (Ref.~\cite{Koe12a}
even carries the title ``Neutron Resonance Data exclude RMT''.)
Indeed, for the Pt isotopes, the analysis rejects agreement with RMT
with a statistical significance of at least 99.997\%
probability~\cite{Koe10}. For the NDE, the corresponding figure is
99.17\%~\cite{Koe11}. Some of these results have found wide
attention~\cite{Rei10, Wei10, Cel11, Vol11} eroding, as they seemingly
do, one of the cornerstones of the statistical theory of nuclear
reactions~\cite{Wei09, Mit10}.

While fully in agreement with the authors of Refs.~\cite{Koe10, Koe11}
concerning the need for a clean separation of $s$-wave and $p$-wave
resonances, we show in this paper that the cutoff procedure actually
used in Ref.~\cite{Koe11} for the analysis of the NDE introduces a
systematic error (this additional error appears to have been included
in Ref.~\cite{Koe10}). Typically, that error is as large as the error
deduced from the width of the maximum of the likelihood function.
Valid conclusions on agreement with RMT predictions can be drawn only
if that error is properly taken into account in the data analysis.

In Section II, we describe the maximum-likelihood method as used in
the analysis of neutron resonance data with a cutoff. Section III
illustrates how the combination of applying a cutoff procedure to the
data and finite-size-of-data effects can produce maximum-likelihood
results that are very different from the underlying data. In Section
IV, we describe simulations that quantify the magnitude of these
effects and demonstrate that the effects are indeed significant for
this particular analysis. Section V provides specific numerical
results for the NDE, and we summarize our results in Section VI.


\section{Maximum-Likelihood Analysis}

For resonances close to threshold (located at energy $E = 0$),
$s$-wave ($p$-wave) neutron widths have an intrinsic energy dependence
$E^{1/2}$ ($E^{3/2}$, respectively), with $E$ taken at the neutron
resonance energy. The transition to reduced widths removes the
$E^{1/2}$ dependence of $s$-wave neutron widths. The remaining linear
energy dependence of $p$-wave resonance widths is used to suppress the
latter with the help of a cutoff first proposed in Ref~\cite{Cam94}.
Cutoffs that depend both linearly~\cite{Koe11} and
non-linearly~\cite{Koe10} on resonance energy have been employed. All
measured neutron widths smaller than the cutoff were removed from the
data set. The distribution of the remaining widths was then analyzed using
a maximum likelihood (ML) method. The method tests for agreement with
the Porter-Thomas distribution (PTD) predicted by RMT. The PTD is a
$\chi^2$-distribution with a single degree of freedom ($\nu = 1$). For
comparison, distribution functions with other values of $\nu$ were
admitted. The test determined which of these gave best agreement with
the data. This led to the above-mentioned rejection of RMT.

The normalized $\chi^2$-distribution for $\nu$ degrees of freedom is
\begin{equation}
g(x, \nu, \langle x \rangle) = \frac{\nu}{2 \Gamma(\nu/2) \langle x
\rangle} \bigg( \frac{\nu x}{2 \langle x \rangle} \bigg)^{(\nu/2) - 1}
\exp \bigg\{ - \frac{\nu x}{2 \langle x \rangle} \bigg\} \ .
\label{1}
\end{equation}
Here $\langle x \rangle$ denotes the average of $x$, and $\Gamma$ is
the complete Gamma function. When the distribution is cut off at
$x_{\rm min} > 0$ the resulting normalized function $g_{\rm cut}$ has
the form $g_{\rm cut}(x, \nu, \langle x \rangle) = (1 /C) g(x, \nu,
\langle x \rangle) \Theta(x - x_{\rm min})$ where $\Theta$ is the
Heaviside function. For an energy-dependent cutoff with value $x_{\rm
  min}(i)$ at resonance energy $E_i$, and with $T_i = x_{\rm min}(i) /
\langle x \rangle$, the cutoff-dependent normalization constant $C_i$
is given by
\begin{equation}
C_i = \frac{\Gamma(\nu/2, \nu T_i / 2)}{\Gamma(\nu / 2)}
\label{2}
\end{equation}
where $\Gamma(x, y)$ is the incomplete Gamma function. The probability
density function for finding a width $x_i > x_{\rm min}$ for a
resonance with energy $E_i$ is then $(1 / C_i) g(x_i, \nu, \langle x
\rangle)$.

When the energy-dependent cutoff procedure is applied to a set of $N$
reduced widths, there remain $N_{\rm incl} \leq N$ included resonances
with energies $E_i$, $i = 1, \ldots, N_{\rm incl}$ and with reduced
widths $x_i > x_{\rm min}(i)$. The likelihood function $L$ is the
product of the corresponding probabilities,
\begin{equation}
L(\nu, \langle x \rangle) = \prod_{i = 1}^{N_{\rm incl}} \frac{1}{C_i}
g(x_i, \nu, \langle x \rangle) \ .
\label{3}
\end{equation}
The maximum of $\ln L$ as a function of $\nu$ and $\langle x \rangle$
determines the most likely values of these parameters. The width of
$\ln L$ in the vicinity of the maximum determines the statistical
significance with which a value of $\nu$ that differs from the value
at the maximum, is rejected. That is the basis for the figures cited
above of $99.17$ per cent for the NDE and of $99.997$ per cent for the
Pt isotopes. (In actual fact the method used in Ref.~\cite{Koe11} is
slightly different. Instead of the product of normalization factors
$C_i$, a single joint normalization factor $C$ to the power $N_{\rm
  incl}$ was used. That factor was defined as a suitable average of
the $C_i$. For the ML analysis with an energy-independent cutoff
studied in the present paper the two methods are identical. Moreover,
the additional error due to the cutoff and explained below seemingly
was taken into account in Ref.~\cite{Koe10}.)


\section{Test of the Cutoff Method}

The reliability of the maximum-likelihood analysis with a cutoff is
adversely affected by (i) the cutoff itself (i.e., even for an
infinite data set) and (ii) by finite-size-of-data effects. We display
these features by using a combination of analytical reasoning and
numerical results. The maximum-likelihood method determines the
extremum of $\ln L(\nu, \langle x \rangle)$ as a function of $\nu$ and
$\langle x \rangle$. At the extremum, we have
\begin{equation}
\frac{\partial}{\partial \nu} \ln L(\nu, \langle x \rangle) = 0 \ , \
\frac{\partial}{\partial \langle x \rangle} \ln L(\nu, \langle x
\rangle) = 0 \ .
\label{4}
\end{equation}
We use an energy-independent cutoff $x_{\rm min}$ throughout. Combining
that with Eqs.~(\ref{1}) and (\ref{2}) we find that Eqs.~(\ref{4})
take the form
\begin{eqnarray}
\label{5}
&& \ln \left( \frac{\nu}{2 } \right) - \ln \langle x \rangle - 2
\frac{\partial \ln \Gamma \left( \frac{\nu}{2} , \frac{\nu x_{min}}{2
\langle x \rangle} \right)} {\partial \nu} \nonumber \\
&& \qquad + \langle \ln x_i \rangle + 1 - \frac{\langle x_i \rangle}
{\langle x \rangle} = 0 \ ,
\end{eqnarray}
\begin{equation}
\langle x \rangle = \langle x_i \rangle + \frac{x_{\rm min}}{\Gamma(x,
y)} \frac{\partial \Gamma(x, y)}{\partial y} \bigg|_{x = (\nu/2), y = T
(\nu/2)} \ .
\label{6}
\end{equation}
Here $T = x_{\rm min} / \langle x \rangle$ and
\begin{eqnarray}
\langle \ln x_i \rangle \equiv \frac{1}{N_{\rm incl}} \sum_{i =
1}^{N_{\rm incl}} \ln x_i \ , \
\langle x_i \rangle \equiv \frac{1}{N_{\rm incl}} \sum_{i = 1}^{N_{\rm incl}}
x_i \ .
\label{7}
\end{eqnarray}
The quantities in Eqs.~(\ref{7}) depend on the actual data set and on
the value chosen for $x_{\rm min}$ (which determines the number
$N_{\rm incl}$ of resonances retained in the analysis). The values for
$\langle x_i \rangle$ and $\langle \ln x_i \rangle$ given by
Eqs.~(\ref{7}) determine $\nu$ and $\langle x \rangle$ as solutions of
Eqs.~(\ref{5}) and (\ref{6}).

We investigate the method by considering Eqs.~(\ref{5}) and (\ref{6})
separately. For fixed input values of $x_{\rm min}$, $\langle x_i
\rangle$, and $\langle \ln x_i \rangle$, each of these two equations
connects $\nu$ with $\langle x \rangle$ and, thus, defines a curve in
the two-dimensional $\nu - \langle x \rangle$ plane. The point of
intersection of these two curves determines the most likely values of
$\nu$ and $\langle x \rangle$. We show how these curves (and their
point of intersection) change with both a change of the cutoff
parameter $x_{\rm min}$ and a change of the number $N$ of resonance widths used in
the analysis. Our input values are $\nu = 1$ and $\langle x \rangle =
1$. If correct, the maximum-likelihood analysis must reproduce these
values.

\begin{figure}
  \includegraphics[width=3.5in]{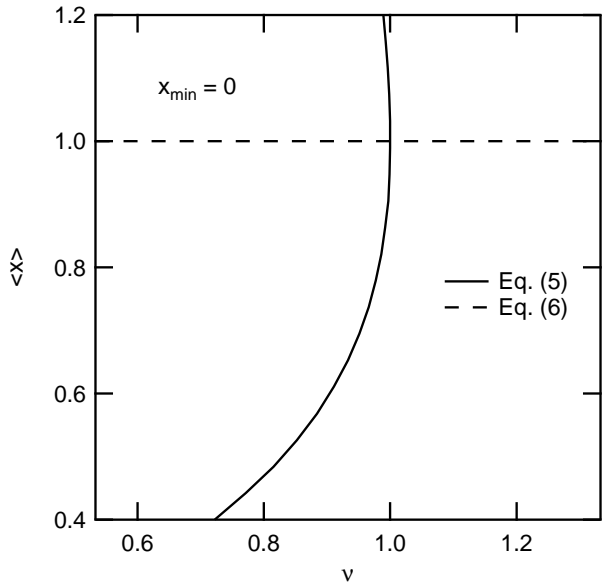}
  \caption{\label{fig1} For $N = \infty$ and $x_{\rm min} = 0$, the
    straight line (the curved line) shows $\langle x \rangle$ versus
    $\nu$ as given by Eq.~(\ref{6}) (by Eq.~(\ref{5}), respectively).
    The point of intersection determines the extremum of the
    maximum-likelihood function.}
\end{figure}

\begin{figure}
\includegraphics[width=3.5in]{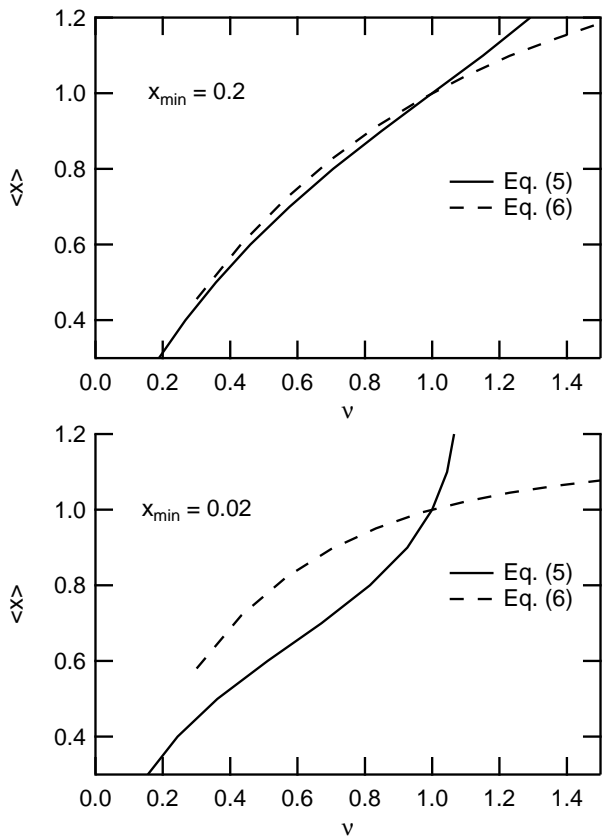}
  \caption{\label{fig2} Same as Figure~\ref{fig1} but for $x_{\rm min}
    = 0.2$ (upper part) and for $x_{\rm min} = 0.02$ (lower part).}
\end{figure}

We begin with the case $N = \infty$ (which implies $N_{\rm incl} =
\infty$). In that case there is no finite-size-of-data error. Given
the Gaussian distribution with unit width of the reduced width
amplitudes $y_i$, both $\langle x_i \rangle$ and $\langle \ln x_i
\rangle$ with $x_i = y^2_i$ can be calculated analytically for every
value of $x_{\rm min}$. For $x_{\rm min} = 0$ we obtain the two curves
shown in Fig.~\ref{fig1}.
The horizontal dashed line corresponds to
Eq.~(\ref{6}), and the solid curve that intersects the dashed line
nearly vertically corresponds to Eq.~(\ref{5}). As expected, the two
curves have a single well-defined point of intersection at $\nu = 1$
and $\langle x \rangle = 1$. Both curves change as $x_{\rm min}$ is
increased. The curve representing Eq.~(\ref{6}) is deformed and
rotated in a counter-clockwise direction while the curve representing
Eq.~(\ref{5}) is likewise deformed but rotated in a clockwise
direction. For $x_{\rm min} = 0.02$ and for $x_{\rm min} = 0.2$, the
results are shown in Fig.~\ref{fig2}.
The two curves still intersect at $\nu = 1$ and $\langle x \rangle =
1$ but become ever more parallel as $x_{\rm min}$ increases. The
effect of this is that any changes in the values of $\langle x_i
\rangle$ or of $\langle \ln x_i \rangle$ that occur for finite values
of $N$ can cause significant shifts in the values of $\nu$ and
$\langle x \rangle$ that maximize the likelihood function $L$.

\begin{figure}[b]
\includegraphics[width=3.5in]{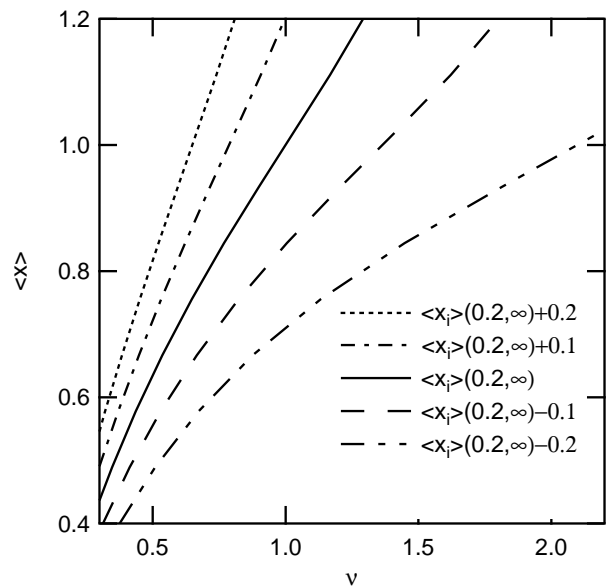}
  \caption{\label{fig3} Dependence of the curve representing
    Eq.~(\ref{5}) on changes of $\langle x_i \rangle$ as given in the
    figure, for a fixed value of $\langle \ln x_i \rangle$ and for $N
    = \infty$, $x_{\rm min} = 0.2$. }
\end{figure}

\begin{figure}
\includegraphics[width=3.5in]{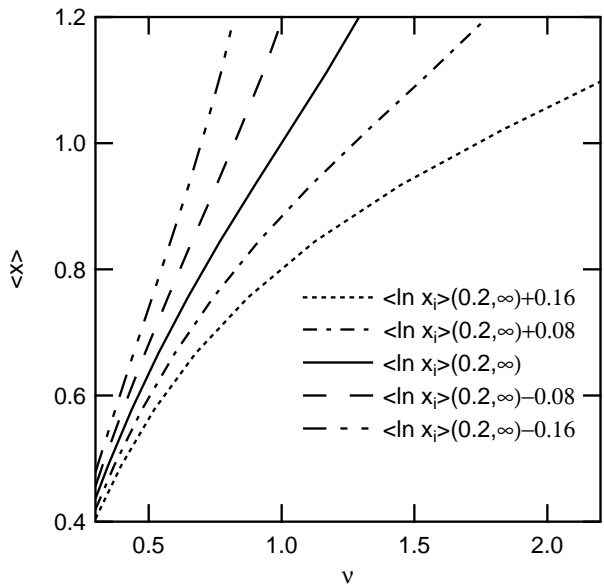}
  \caption{\label{fig4} Dependence of the curve representing
    Eq.~(\ref{5}) on changes of $\langle \ln x_i \rangle$ as given in the
    figure, for a fixed value of $\langle x_i \rangle$ and for $N
    = \infty$, $x_{\rm min} = 0.2$. }
\end{figure}

The fluctuations in $\langle x_i \rangle$ or $\langle \ln x_i \rangle$
for spectra of finite size $N$ are expected to be of order $1 /
\sqrt{N}$. For $N = 100$, a reasonable spectrum size for nuclear data,
that amounts to $10 \%$. In Figs.~\ref{fig3} and \ref{fig4} we
illustrate the effect such fluctuations might have by changing in
Eq.~(\ref{5})  (somewhat unrealistically) $\langle x_i
\rangle$ and $\langle \ln x_i \rangle$ separately and independently.
For $N = \infty$ and $x_{\rm min} = 0.2$, Fig.~\ref{fig3} shows how
the curve representing Eq.~(\ref{5}) is
changed when $\langle x_i \rangle$ is varied while $\langle \ln x_i
\rangle$ is held fixed. Fig.~\ref{fig4} shows the same for fixed
$\langle x_i \rangle$ and a variation of $\langle \ln x_i \rangle$. We
note that an increase in $\langle x_i \rangle$ causes the curves to
shift in one direction while an increase in $\langle \ln x_i \rangle$
causes them to shift in the other direction. We see that as a result
of the finite-size-of data-errors in $\langle x_i \rangle$ and
$\langle \ln x_i \rangle$, the curves change significantly, and so do
(as it turns out) their points of intersection. The dependence of the
curve representing Eq.~(\ref{6}) on changes of $\langle x_i \rangle$
is similar to that shown in Fig.~\ref{fig3}.

\section{Simulations}

\begin{figure}
\includegraphics[width=3.5in]{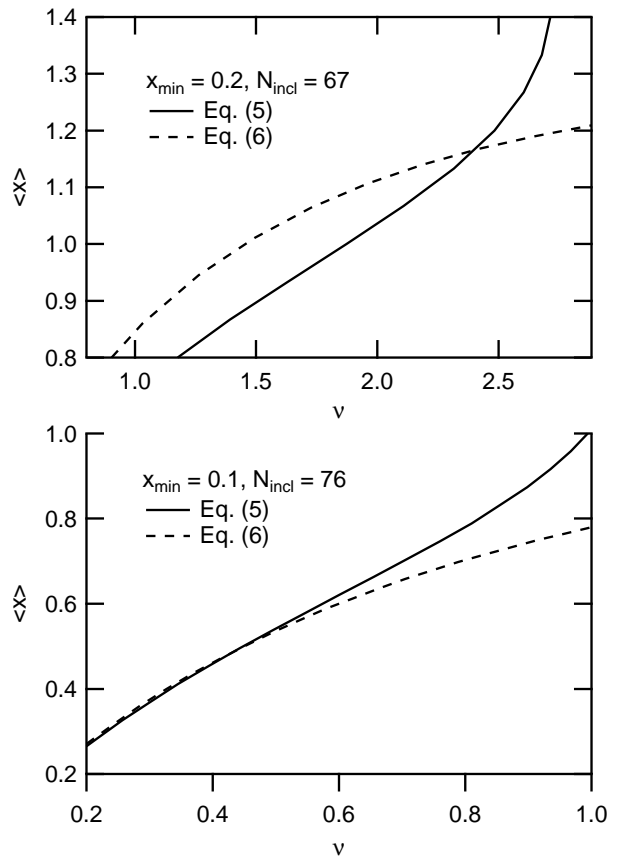}
  \caption{\label{fig5} Curves representing Eqs.~(\ref{5}) and
    (\ref{6}) for two individual random drawings of $N=100$ widths and
    for cutoff values $x_{\rm min} = 0.2$ (upper figure) and $x_{\rm
      min} = 0.1$ (lower figure). For these two cases, the points of
    intersection (i.e., the values obtained by the ML method) are at
    $\nu \approx 2.40$, $\langle x \rangle \approx 1.16$ (upper
    figure) and $\nu \approx 0.44$, $\langle x \rangle \approx 0.50$
    (lower figure).}
\end{figure}

We have shown that applying a cutoff as described in
Refs.~\cite{Cam94} and \cite{Koe11} has the potential to shift
significantly the ML values of $\nu$ and $\langle x \rangle$. We now
show that significant shifts of $\nu$ do actually occur in finite data
sets. We have simulated finite values of $N$ with the help of a
random-number generator. We have drawn $N$ real numbers $y_i$, $i = 1,
\ldots, N$ from a Gaussian probability distribution with unit width.
The squares determine $N$ widths $x_i = y^2_i$. By construction, in
the limit $N \to \infty$ these obey the PTD with $\langle x \rangle =
1$. Each of these widths is associated with one of $N$ resonances. We
have repeated that procedure $2500$ times, thereby generating for each
value of $N$ an ensemble of widths. For every member of the ensemble,
we use the same cutoff $x_{\rm min}$ and determine $N_{\rm incl}$,
$\langle x_i \rangle$ and $\langle \ln x_i \rangle$. These values are
used to construct the curves representing Eqs.~(\ref{5}) and
(\ref{6}). As a general rule we find that for finite $N$ and with
increasing $x_{\rm min}$, the curve representing Eq.~(\ref{5})
(Eq.~(\ref{6})) is rotated in the clockwise direction (in the
counter-clockwise direction), as in the case for $N = \infty$. In
addition, the shapes of the two curves are tilted in a way that
depends on the specific $N$ values of the widths $x_i$.
Fig.~\ref{fig5} shows two cases for $N = 100$, a value relevant to the
sets of neutron widths. In both cases, the rotation of the two curves
with increasing $x_{\rm min}$ is clearly seen. In one case, the point
of intersection of the two curves representing Eqs.~(\ref{5}) and
(\ref{6}) occurs for $\nu$ significantly less than unity; for the
other case, the point of intersection produces a value of $\nu$
noticeably larger than unity.

\begin{figure}
\includegraphics[width=3.5in]{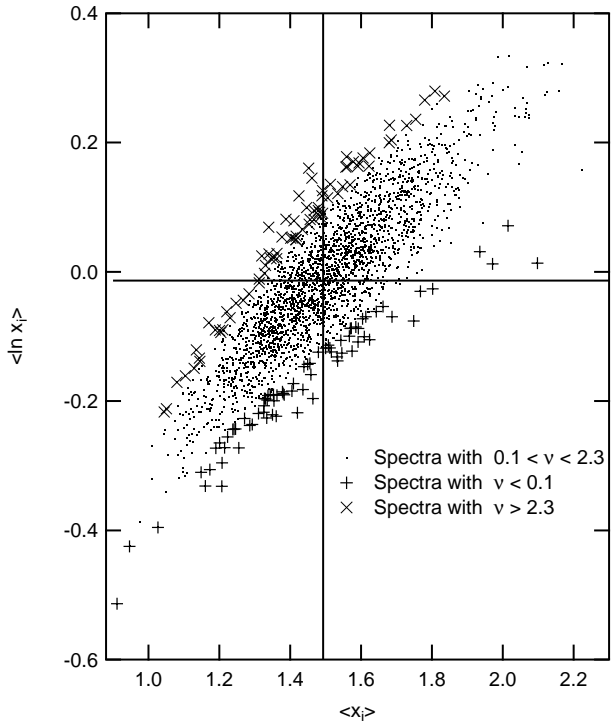}
  \caption{\label{fig6} Scatter plot of the ensemble of 2500 values of
    $\langle x_i \rangle$ and $\langle \ln x_i \rangle$ obtained from
    random drawings of $N = 100$ widths with a cutoff $x_{\rm min}
    = 0.2$. The crossing point of the vertical and the horizontal
    lines indicates the solution for $N = \infty$. The crosses
    correspond to solutions with $\nu > 2.3$, and the plus signs
    correspond to solutions with $\nu < 0.1$. The dots represent the
    remaining data.}
\end{figure}

Fig.~\ref{fig6} shows a scatter plot of
the values of $\langle x_i \rangle$ and $\langle \ln x_i \rangle$ over
the ensemble of $2500$ random drawings of $N = 100$ widths when a
cutoff $x_{\rm min} = 0.2$ is employed. We note that the fluctuations
used in Figs.~\ref{fig3} and \ref{fig4} are realistic. Cases where
values $\nu < 0.1$ or $\nu > 2.3$ are obtained as solutions of
Eqs.~(\ref{5}) and (\ref{6}) are indicated separately in the
figure. For every value of $\langle x_i \rangle$ the highest (lowest)
values of $\nu$ occur for the highest (lowest) values of $\langle \ln
x_i \rangle$. The values $\nu < 0.1$ and $\nu > 2.3$ are chosen so as
to obtain approximately similar numbers of extreme cases.

\begin{figure}
\includegraphics[width=3.5in]{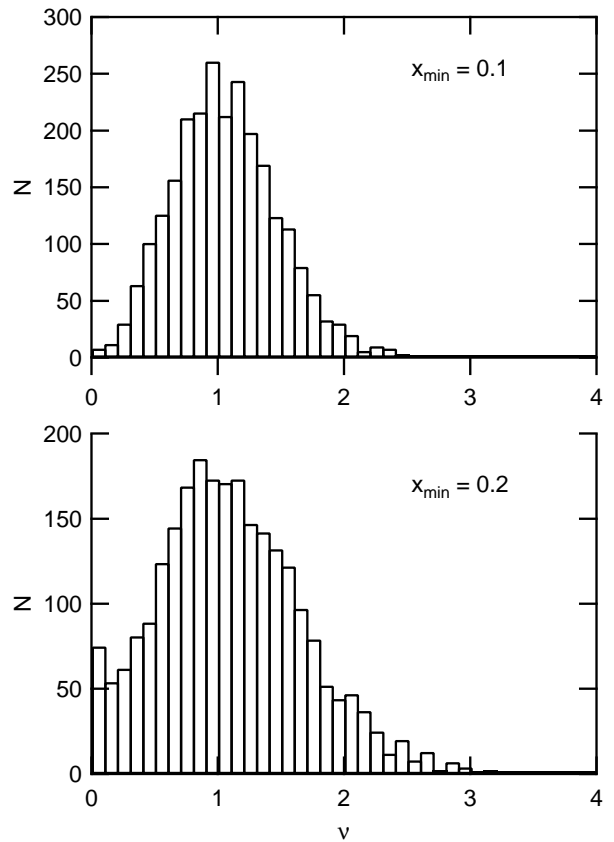}
  \caption{\label{fig7} Distributions of $\nu$ values obtained as
    solutions of Eqs.~(\ref{5}) and (\ref{6}) for an ensemble of $2500$
    sets of $N = 100$ randomly chosen widths with cutoffs of $x_{\rm
      min} = 0.2$ (lower figure) and $x_{\rm min} = 0.1$. (upper figure). }
\end{figure}

As a summary of these results, Fig.~\ref{fig7} shows the distribution
of $\nu$ values that are obtained as solutions of Eqs.~(\ref{5}) and
(\ref{6}) for the 2500 random drawings of $N = 100$ widths with cutoff
values of $x_{\rm min} = 0.1$ and $x_{\rm min} = 0.2$. The corresponding
means and standard deviations of these distributions are 1.06 and 0.42
for $x_{\rm min} = 0.1$ and 1.10 and 0.59 for $x_{\rm min} = 0.2$.
This illustrates both a small bias toward increasing values of $\nu$
obtained with this method (a fact which was mentioned briefly by
Camarda \cite{Cam94}) as well as a significant increase in the
standard deviations as the cutoff parameter $x_{min}$ is increased.

\section{Applications}

We now apply this method to the NDE. In each case, we estimate
$\sigma_c$, the uncertainty associated with the combination of the
finite-size-of-data effect and the application of a cutoff, by
utilizing the simulation process described in Section IV: 2500 sets of
widths, each set of size $N$, are sampled from a Porter-Thomas
distribution. A cutoff varying linearly with energy is then applied,
as described in Ref.~\cite{Koe11}; the cutoff is parameterized by a
quantity $T_{max}$, where $x_{min} = T_{max} \, E_n/E_{max}$ ($E_n$ is
the neutron energy of the resonance in question, and $E_{max}$ is the
maximum energy in that set of widths). The values of $N$ and $T_{max}$ for each
member of the NDE  are taken from Ref.~\cite{Koe11}. We
assume equally spaced resonances in our simulations. It is known that
neutron resonance energy levels show significant short- and long-range
correlations \cite{Haq82,Boh83,Boh85}. Because of the ensuing stiffness of the spectrum,
this
assumption of equal spacings should have minimal impact on our results. We take
$\sigma_c$ to be the half-width of the central 68\% of the
distribution of the parameter $\nu$. The overall uncertainty is then
obtained by combining $\sigma_c$ in quadrature with the uncertainty
$\sigma_{ML}$ given by the previous maximum-likelihood analysis.

Table \ref{NDEtab} summarizes the results. The first five columns are
taken from Ref.~\cite{Koe11}. The next column shows $\sigma_c$. The
last column shows the total uncertainty when $\sigma_c$ is included.

\renewcommand{\arraystretch}{1.5}
\begin{table}
\caption{\label{NDEtab} Results for the data sets included in the
  Nuclear Data Ensemble. The first three columns summarize the data.
  The next two show the cutoff parameter $T_{max}$ used by
  Koehler~\cite{Koe11} in his analysis, the resulting
  maximum-likelihood estimate for $\nu$, and the uncertainty
  $\sigma_{ML}$. The last two columns show the values of $\sigma_c$
  and the resulting overall uncertainties resulting from the
  application of a cutoff that is linear in energy, as described in
  the text.}
\begin{ruledtabular}
\begin{tabular}{ccccccc}
Nuclide & $N$ & $E_{max}$ & $T_{max}$
 & $\nu$ & $\sigma_c$ & $\nu$ \\
& & (keV) & & Ref \cite{Koe11} & &
This Work \\
\colrule
$^{64}$Zn & 103 & 367.55 & 0.05 & 1.54$^{+0.29}_{-0.26}$ & 0.22 & 1.54$^{+0.36}_{-0.34}$ \\
$^{66}$Zn & 65 & 297.63 & 0.05 & 0.74$^{+0.27}_{-0.25}$ & 0.30 & 0.74$^{+0.40}_{-0.39}$ \\
$^{68}$Zn & 45 & 247.20 & 0.05 & 0.95$^{+0.36}_{-0.32}$ & 0.36 & 0.95$^{+0.51}_{-0.48}$ \\
$^{114}$Cd & 17 & 3.3336 & 0.45 & 2.0$^{+1.5}_{-1.2}$ & 1.3 & 2.0$^{+2.0}_{-1.8}$ \\
$^{152}$Sm & 70 & 3.665 & 0.1 & 1.55$^{+0.40}_{-0.38}$ & 0.33 & 1.55$^{+0.52}_{-0.50}$ \\
$^{154}$Sm & 27 & 3.0468 & 0.1 & 1.32$^{+0.65}_{-0.55}$ & 0.57 & 1.32$^{+0.86}_{-0.79}$ \\
$^{154}$Gd & 19 & 0.2692 & 0.2 & 0.49$^{+0.64}_{-0.48}$ & 0.91 & 0.49$^{+1.11}_{-1.03}$ \\
$^{156}$Gd & 54 & 1.9908 & 0.2 & 1.44$^{+0.51}_{-0.49}$ & 0.46 & 1.44$^{+0.69}_{-0.67}$ \\
$^{158}$Gd & 47 & 3.9827 & 0.2 & 1.17$^{+0.54}_{-0.47}$ & 0.49 & 1.17$^{+0.73}_{-0.68}$ \\
$^{160}$Gd & 21 & 3.9316 & 0.2 & 0.83$^{+0.75}_{-0.65}$ & 0.80 & 0.83$^{+1.10}_{-1.03}$ \\
$^{160}$Dy & 18 & 0.4301 & 0.2 & 1.41$^{+1.0}_{-0.83}$ & 0.90 & 1.41$^{+1.34}_{-1.22}$ \\
$^{162}$Dy & 46 & 2.9572 & 0.2 & 0.99$^{+0.47}_{-0.43}$ & 0.48 & 0.99$^{+0.67}_{-0.64}$ \\
$^{164}$Dy & 20 & 2.9687 & 0.2 & 2.3$^{+1.2}_{-1.0}$ & 0.81 & 2.3$^{+1.4}_{-1.3}$ \\
$^{166}$Er & 109 & 4.1693 & 0.3 & 1.85$^{+0.49}_{-0.45}$ & 0.35 & 1.85$^{+0.60}_{-0.57}$ \\
$^{168}$Er & 48 & 4.6711 & 0.3 & 1.32$^{+0.62}_{-0.55}$ & 0.56 & 1.32$^{+0.84}_{-0.78}$ \\
$^{170}$Er & 31 & 4.7151 & 0.3 & 3.6$^{+1.6}_{-1.3}$ & 0.71 & 3.6$^{+1.8}_{-1.5}$ \\
$^{172}$Yb & 55 & 3.9000 & 0.06 & 0.70$^{+0.30}_{-0.26}$ & 0.34 & 0.70$^{+0.45}_{-0.43}$ \\
$^{174}$Yb & 19 & 3.2877 & 0.06 & 1.29$^{+0.68}_{-0.58}$ & 0.62 & 1.29$^{+0.92}_{-0.85}$ \\
$^{176}$Yb & 23 & 3.9723 & 0.06 & 1.05$^{+0.65}_{-0.55}$ & 0.53 & 1.05$^{+0.84}_{-0.76}$ \\
$^{182}$W & 40 & 2.6071 & 0.15 & 1.50$^{+0.62}_{-0.55}$ & 0.51 & 1.50$^{+0.80}_{-0.75}$ \\
$^{184}$W & 30 & 2.6208 & 0.15 & 0.99$^{+0.54}_{-0.48}$ & 0.61 & 0.99$^{+0.81}_{-0.48}$ \\
$^{186}$W & 14 & 1.1871 & 0.15 & 1.32$^{+0.93}_{-0.75}$ & 1.0 & 1.3$^{+1.4}_{-1.3}$ \\
$^{232}$Th & 178 & 2.988 & 0.26 & 1.78$^{+0.36}_{-0.34}$ & 0.25 & 1.78$^{+0.44}_{-0.42}$ \\
$^{238}$U & 146 & 3.0151 & 0.47 & 1.02$^{+0.39}_{-0.34}$ & 0.33 & 1.02$^{+0.51}_{-0.47}$ \\
\end{tabular}
\end{ruledtabular}
\end{table}
\renewcommand{\arraystretch}{1.}

In general, $\sigma_c$ in each case has approximately the same
magnitude as the value of $\sigma_{ML}$ for that nuclide. The
resulting weighted average of $\nu$ for the NDE when the cutoff error
is included is $1.25 \pm 0.13$, as compared to the value $1.217 \pm
0.092$ listed in Ref.~\cite{Koe11}. Even though the overall
uncertainty is larger when $\sigma_c$ is included in the analysis, the
weighted average of $\nu$ has also increased so that the level of
significance changes very little from the figure given in
Ref.~\cite{Koe11}. Most of the increase occurs because the overall
uncertainty for the largest data set, $^{232}$Th, is increased by a
relatively smaller amount than the other data sets, and the value of
$\nu$ for these data is rather high. To put this in perspective, the
weighted average of $\nu$ when $^{232}$Th is not included is $1.20 \pm
0.13$, a value which does not show nearly as significant a difference
from the RMT expectation of $\nu = 1$.

\section{Conclusions}

It has been known for a long time that various eigenvalue statistics
(such as $\Delta_3$~\cite{Dys63}) are extremely sensitive to missing
or spurious levels. The effects of spurious or missing levels on the
reduced-width distribution have not been nearly as well studied.
Koehler~\cite{Koe11} has analyzed the reduced neutron widths in the
nuclear data ensemble using a cutoff method to remove the smallest
reduced widths, many of which probably belong to $p$-wave resonances.
Unfortunately, the procedure also carries the risk of removing small
$s$-wave resonances that should be included in the data set.

We have studied in detail how applying a constant cutoff affects the
maximum-likelihood analysis of sets of theoretical and of
computer-generated neutron widths generated from a PTD. Expressing the
condition for the maximum of the likelihood function $\ln L$ in terms
of two equations, we have investigated numerically the dependence of
these two equations on the cutoff $x_{\rm min}$ and on the number $N$
of resonance widths. We have found it convenient to represent the two
equations graphically, their point of intersection giving the values
of $\nu$ and $\langle x \rangle$ that correspond to the maximum of
$\ln L$. We have focused on a realistic cutoff value $x_{\rm min} =
0.2$ amounting to $20$ per cent of the mean value of all widths. For
$N = \infty$, i.e., in the absence of finite-size-of-data-errors, the
cutoff rotates the two curves in such a way that they run almost
parallel. Therefore, any changes in the values of $\langle x_i
\rangle$ or of $\langle \ln x_i \rangle$ that occur for finite values
of $N$ can cause significant shifts in the values of $\nu$ and
$\langle x \rangle$ that maximize the likelihood function $L$. The
rotation persists for finite values of $N$ but in addition each of the
two curves is deformed, causing the wide scatter of solutions $\nu$
shown in Fig.~\ref{fig7} for realistic cutoff values $0.1$ and $0.2$.
The full width at half maximum of both distributions shown in
Fig.~\ref{fig7} is about unity. We conclude that in realistic cases
systematic errors (due to the cutoff) and finite-size-of-data errors
(due to the finite number $N$ of resonance widths experimentally
available) combine in such a way that the resulting error on $\nu$ is
of order unity. We refer to the combination of both errors as the
cutoff error.

The implications of the cutoff error for the NDE are displayed in
Table~\ref{NDEtab}. Compared to the analysis of Ref.~\cite{Koe11} the
overall error is significantly increased for every nucleus. The
weighted average over all nuclei gives $\nu = 1.25 \pm 0.13$. As
expected, the total error is larger than that obtained in
Ref.~\cite{Koe11}. However, the mean value of $\nu$ is also increased
in such a way that the significance of the deviation of $\nu$ from the
RMT value remains nearly the same. We conclude that inclusion of the
cutoff error does not remove the strong discrepancy (first displayed
in Ref.~\cite{Koe11}) between the distribution of neutron widths as
predicted by RMT and the NDE. It must be borne in mind, however, that
the strong discrepancy is essentially caused by the deviation from the
PTD found for $^{232}$Th.

\begin{acknowledgments}

One of us (JFS) thanks P. E. Koehler for very helpful discussions
concerning the application of the maximum-likelihood method. We are
grateful to an anonymous referee who pointed out an essential error in
the original version of this paper~\cite{Shr12}. One of us (GEM)
acknowledges the support of the US Department of Energy via grants
No. DE-FG52-09NA29460 and No. DE-FG02-97-ER41042.

\end{acknowledgments}


\end{document}